\documentclass[
superscriptaddress,
twocolumn,
amsmath,amssymb,nofootinbib
]{revtex4}

\usepackage{graphicx,color}
\usepackage{dcolumn}
\usepackage[final]{pdfpages}

\begin{document}
\title{Collective ECM remodeling organizes 3D collective cancer invasion}

\author{Jihan Kim\footnote{These authors contributed equally to this work.}}
\affiliation{Department of Physics, Oregon State University, 301
  Weniger Hall, Corvallis, OR, USA, 97331-6507}
\author{Yu Zheng\footnotemark[1]}
\affiliation{Department of Physics, Arizona State University,
Tempe, AZ 85287-6106}
\author{Amani A. Alobaidi\footnotemark[1]}
\affiliation{Department of Physics, Oregon State University, 301
  Weniger Hall, Corvallis, OR, USA, 97331-6507}
\author{Hanqing Nan}
\affiliation{Materials Science and Engineering, Arizona State
University, Tempe, AZ 85287-6106}
\author{Jianxiang Tian}
\affiliation{Department of Physics, Qufu Normal University, Qufu
273165, China} \affiliation{Materials Science and Engineering,
Arizona State University, Tempe, AZ 85287-6106}
\author{Yang Jiao}
\email[correspondence sent to: ]{yang.jiao.2@asu.edu}
\affiliation{Materials Science and Engineering, Arizona State
University, Tempe, AZ 85287-6106} \affiliation{Department of
Physics, Arizona State University, Tempe, AZ 85287-6106}
\author{Bo Sun}
\email[correspondence sent to: ]{sunb@onid.orst.edu}
\affiliation{Department of Physics, Oregon State University, 301
  Weniger Hall, Corvallis, OR, USA, 97331-6507}

\begin{abstract}

Tumor metastasis, traditionally considered as
random spreading of individual cancer cells, has been shown to
also involve coordinated collective invasion in the tissue space.
Here we demonstrate experimentally and computationally that
physical interactions between cells and extracellular matrix (ECM)
support coordinated dissemination of cells from tumor organoids.
We find that collective remodeling of the ECM fibrous structure by
cell-generated forces produces reciprocal cues to bias cell
motility. As a result, dissemination of cells from tumor organoids
are controlled by the organoid geometry. Our results indicate that
mechanics and geometry are closely coupled in multicellular
processes during metastasis and morphogenesis. This study also
suggests that migrating cells in 3D ECM represent a distinct class
of active particle system where the collective dynamics is
governed by the micromechanical remodeling of the environment
rather than direct particle-particle interactions.

\end{abstract}
\maketitle
Collective invasion of cancer cells plays an important role in the lethal metastasis of tumors \cite{Segall_collective_review, Pascal2017_collectivereview}. However, the mechanisms that coordinate 3D collective cell motility is not fully understood. A number of mechanisms have been shown to coordinate multicellular invasion. For instance, cell-cell adhesion leads to group invasion as cell clusters \cite{Ewald2013collectiveinvasion,Maheswaran2014}. Strands of cells may follow each other along microtracks created by leader cells \cite{Friedl2007_individualtocollective,Sahai2007_fibroblastcancer}. Communications mediated by diffusive factors have also been shown to coordinate the collective motility of tumor cells by establishing leader-follower phenotypes \cite{Marcus2017}. 

Here we propose and test a distinct mechanism for 3D collective
invasion that only depends on the physical interactions between
cancer cells and their ECM (extracellular matrix). We show both
experimentally and computationally that the reconfigurability of
ECM, as well as the force generation and force sensation of cancer
cells maintain a mechanical conversation between invading tumor
cells in 3D tissue space. As a result, the collective invasion is
coordinated by collective micromechanical remodeling of the ECM,
which leads to organoid-geometry dependent dissemination of the
tumor cells. Our results suggest that ECM-mediated physical
interactions between invasive cells may play a powerful role in
determining the metastatic potential of malignant tumors.


\begin{figure*}[ht]
\centering \includegraphics[width=1.5\columnwidth]{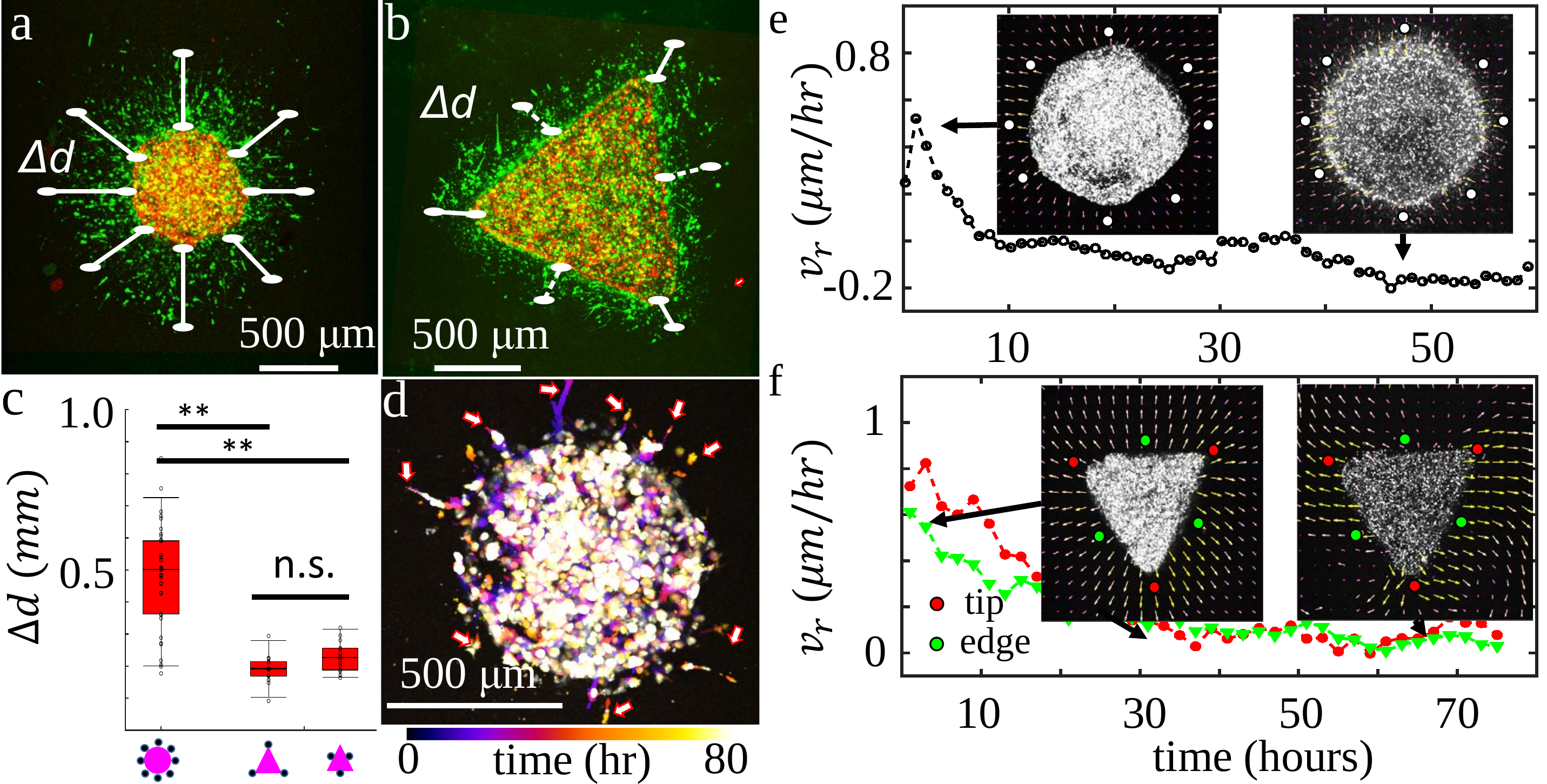}
\caption{Geometry controls tumor organoid invasiveness and ECM remodeling. (a-b) Tumor organoid morphology shown by maximum projection at day 0 (red) and day 10
  (green). For a circular tumor the invasion depth $\Delta d$ is measured along eight
  equally spaced angles (a). For a triangular tumor $\Delta d$ is measured
  along tip and edge directions separately (b). (c) The invasion depths of
  circular and triangular tumors. $N=$5 tumors were measured for each geometry. Statistical comparison are done with t-test. **: p<0.01. n. s. : not significant. (d) Time-coded image projection showing the invasion process
  of a circular MDA-MB-231 organoid. The cells disseminate
  individually (arrows) with little contact after leaving the organoid. (e-f) The average radial velocity of the ECM deformation near
  expanding tumors. In (e) the velocity is averaged over the 8 dotted
  locations shown in the inset. In (f) the velocity is averaged over
  the dotted locations along the edge and tip directions respectively. Insets in (e-f) show the net deformation of the ECM at two time points (black arrows).}\label{fig_invasion_depth}
\end{figure*}

{\bf Geometry controls tumor organoid invasiveness and collective
micromechanical remodeling of ECM.} To study the 3D collective
invasion of cancer cells, we use tumor organoid models created
with DIGME technique \cite{Sun2016DIGME}. Each organoid consists
of approximately 1000 GFP-labeled MDA-MB-231 cells molded into
various shapes in 3D type I collagen ECM (Fig. S1). Confocal
imaging starts immediately after the gelation process completes
(time zero).

We first compare tumors of two different cross-sectional
geometries: circular and triangular. In particular, we measure the
invasion depths $\Delta d$ as the distance between the outer
boundaries of the tumors at day 0 and day 10 as shown in Fig.
\ref{fig_invasion_depth}(a-b). The circular tumors disseminate
rapidly, with the mean invasion depth to be more than 500 $\mu$m.
The triangular tumors, on the other hand, disseminate much slower.
The invasion depths measured in the edge and tip directions are
both around 200 $\mu$m, less than half of the mean invasion depth
of circular tumors (Fig. \ref{fig_invasion_depth}c).

The cells in the organoids are harvested from the same subculture,
and the invasion assays are maintained in the same tissue culture
conditions, it is very unlikely that circular tumors and
triangular tumors consists of two distinct phenotypes. Ruling out
single-cell sources, Fig. \ref{fig_invasion_depth}c suggests the
3D invasion of a tumor organoid is a complex multicellular
phenomenon, where cell-cell and cell-ECM interactions may play
important roles.

We first note that there are rarely any cell-cell adhesions during
the invasion process due to the highly mesenchymal nature of
MDA-MB-231 cells. In distinction from cells exhibiting epithelial
features, MDA-MB-231 cells characteristically disseminate as
isolated individuals (Fig. \ref{fig_invasion_depth}d).
We then consider the possibility of indirect cell-cell
interactions. Previously we reported that the traction forces from
cell pairs create aligned collagen fiber bundles, such that the
remodeled ECM provides a potential channel to mediate mechanical
communications between cancer cells \cite{Sun2017ECMplasticity}.
For an disseminating tumor organoid, the micro-mechanical
remodeling of the ECM could be even stronger due to additive
effects. Therefore, we hypothesize that collective cell remodeling
of the ECM renders the invasion of tumor organoids to depend on
organoid geometry.

To test this hypothesis, we first examine if the ECM deformation
caused by tumor-generated mechanical forces depends on the tumor
geometry. We embed 1-$\mu$m diameter fluorescently-labeled
polystyrene particles in the collagen ECM and use particle image
velocimetry (PIV) to quantify the ECM deformation field. Fig.
\ref{fig_invasion_depth} (e-f) show the radial velocity $v_{r}$ of
the ECM by averaging over symmetric locations (dotted points in
the insets of Fig. \ref{fig_invasion_depth} (e-f)). A circular
tumor first pushes out the ECM ($v_{r}>0$) due to cell spreading
upon seeding (leading to overall expansion of the organoid), then
pulls in the ECM ($v_{r}<0$) with their traction force. As a
result, after 2 days the ECM shows net inward deformation while
some of the cells have already been disseminated from the tumor
(Fig. \ref{fig_invasion_depth}e insets). On the other hand, we
find that a triangular tumor mostly pushes out the ECM, albeit
with a diminishing rate. As a result, the ECM surrounding the
triangular tumor maintains a radially outward net deformation
(Fig. \ref{fig_invasion_depth}f).

{\bf Computational modeling of collective micromechanical
remodeling of ECM and tumor invasion.} To gain further insights on
how collective ECM remodeling modulates the collective cancer
invasion, we devise a multi-scale computational model that takes
into account the fibrous microstructure of the ECM
\cite{Sun2014gel}, nonlinear ECM mechanics
\cite{Fabry2016_3dtractionforce}, as well as cell motility
directed by contact guidance cues \cite{han2016oriented,Levine2018durotaxis}.


Based on the experimental observations, we consider the
dissemination of tumor organoids to start from an expansion phase,
where cells spread and push out the ECM. This is followed by an
invasion phase, where cells pull in the ECM and migrate. We model
tumor cells as polarized active particles with coupled force
generation and locomotion \cite{han2016oriented} (Fig. S2-S7). Cells deform the ECM fibers in their
vicinity, which in turn alters the migration and polarization of
the cells. Explicitly accounting for the reciprocal interactions
between cells and ECM allows us to investigate the collective
migration regulated by the non-local mechanical dialogues among
the cells mediated by the ECM.

\begin{figure}[ht]
\centering
\includegraphics[width=0.99\columnwidth]{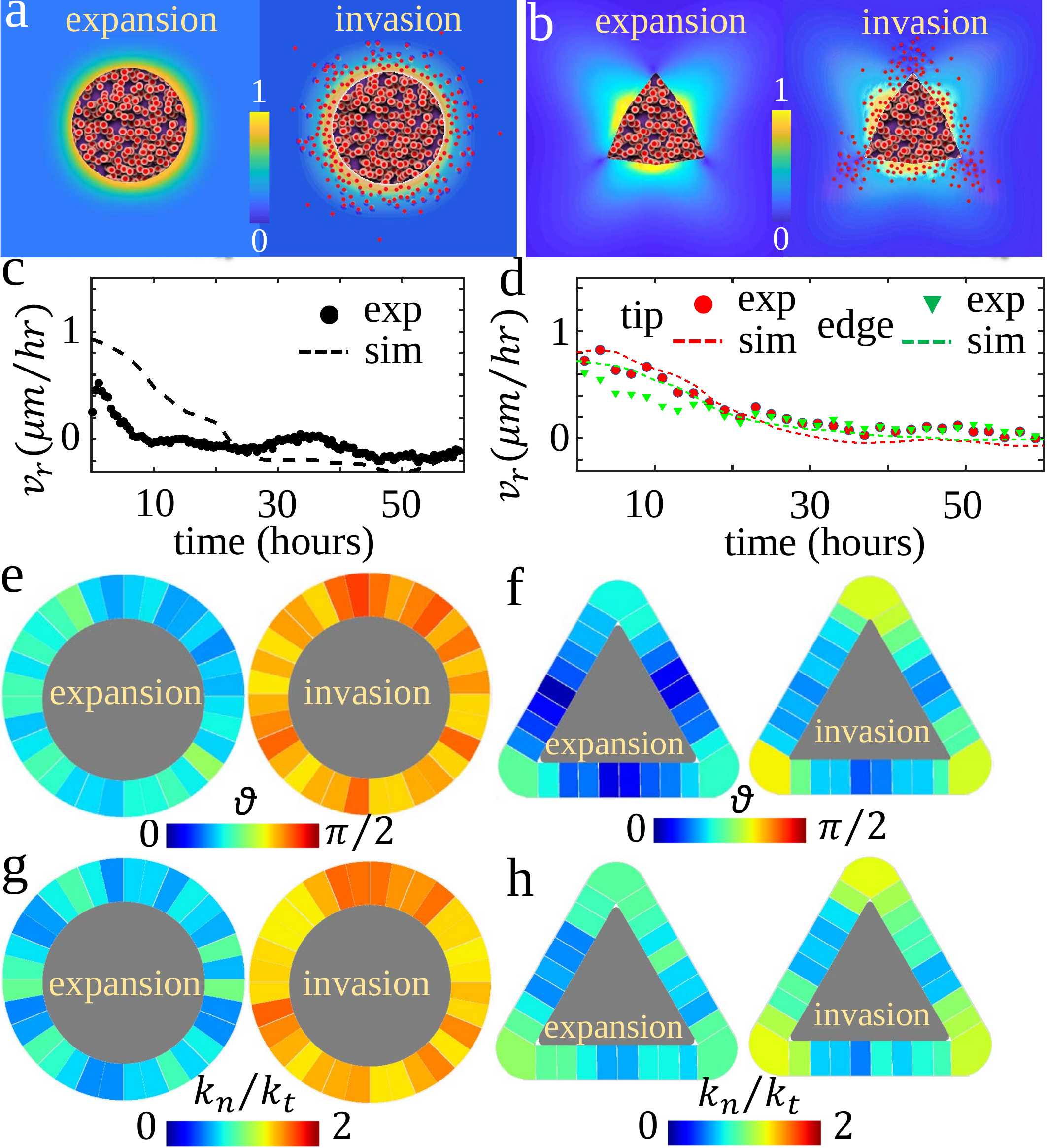}
\caption{Simulated ECM remodeling by circular and triangular tumor
organoids. (a-b) The magnitudes of ECM
displacement fields at 5 hours (representing the expansion phase) and at 50 hours (representing invasion phase) of circular and
triangular organoids. Maximum deformation is normalized to 1. (c-d) Radial velocities of the ECM sampled near circular and triangular organoids at locations corresponding to the experimental measurements in Fig. \ref{fig_invasion_depth}(e-f). Abbreviations: exp: experiment; sim: simulation. (e-f) The orientation of ECM fibers near the organoids at 5 hours (expansion) and at 50 hours (invasion). $\theta$ is measured as the average angle of fibers with respect to their local tangential direction of the tumor boundary. (e-f) The micromechanical anisotropy defined as $k_n/k_t$ at 5 hours (expansion) and at 50 hours (invasion). $k_n$ and $k_t$ are the micromechanical stiffness along the normal direction and tangential direction of the tumor boundary respectively. See Fig. S8 for more details. In (e-h) we sample 100 $\mu$m$^3$ volume elements next to the organoids to obtain the average fiber orientations and stiffness ratios. }\label{fig_simuECM}
\end{figure}

We first employ the computational model to simulate and calculate the
remodeling of the ECM surrounding circular and triangular tumor organoids. Fig. \ref{fig_simuECM} (a-b) show the relative magnitudes of displacement
fields in the ECM, where the maximum value is normalized to 1. Here we measure the ECM properties at 5 hours and 50 hours, two time points empirically determined to represent the expansion and invasion phases respectively.  The magnitude of displacement decays roughly as $1/r^3$ as one moves away from the organoids. After the expansion phase, individual cells 
migrate away from the original organoid. The right panels of Fig. \ref{fig_simuECM}(a-b) show locations of the invaded cells in a typical simulation run. Cell invasion is accompanied with continuous ECM deformation. As shown in Fig.
\ref{fig_simuECM}(c-d), our simulated ECM radial velocity agree well (qualitatively for circular organoids and quantitatively for triangular organoids) with experimental measurements. Importantly, the
contractile deformation is much more pronounced near circular
tumors compared with triangular tumors.

To better reveal the structural remodeling of the ECM, we have
calculated the average orientation of the ECM fibers with respect
to the tangential direction of the tumor surfaces (Fig.
\ref{fig_simuECM}e-f). During the expansion phase, organoids
push the fibers to be aligned parallel to the tumor boundary,
which bias the cell polarization accordingly. Later on, the
pulling forces from the cells reorient the fibers. For circular
tumors, the collective cellular traction force is sufficient to
align the ECM fibers radially (Fig. \ref{fig_simuECM}e),
contributing to the accelerated the dissemination. In contrast,
the fibers remain tangentially aligned along the flat edges of the
triangular tumors (Fig. \ref{fig_simuECM}f). 

The structural remodeling of the ECM significantly reconfigures the micromechanics of the ECM. We find that for both circular and triangular tumors ECM is consistently stiffer in the direction of fiber alignment, and softer in the direction perpendicular to the fibers. Such mechanical cues may further regulate the dynamics and functions of cells through mechanosensing pathways \cite{Robinson2013_mechanosensing}.


Our simulations show that the circular tumors exhibit larger
invasion depths compared with the triangular tumors (Fig. S9), which
is consistent with the experimental observations. To further
explore the mechanisms behind the geometry-dependent collective
migration, we modified the simulation parameters to steer the
cell-ECM interactions. We find that when the cell polarization and
ECM fiber orientations become uncorrelated, the invasion depth of
tumors drastically reduces (Fig. S10). This is consistent
with the experimental results that reducing the level of contact guidance diminished the advantage of circular tumors in
dissemination (Fig. S11). We also find that reducing
the cellular traction forces leads to weaker ECM remodeling, and
thus weaker dependence of tumor invasion dynamics on the organoid
geometry (Fig. S12).

\begin{figure*}[ht]
\centering
\includegraphics[width=1.5\columnwidth]{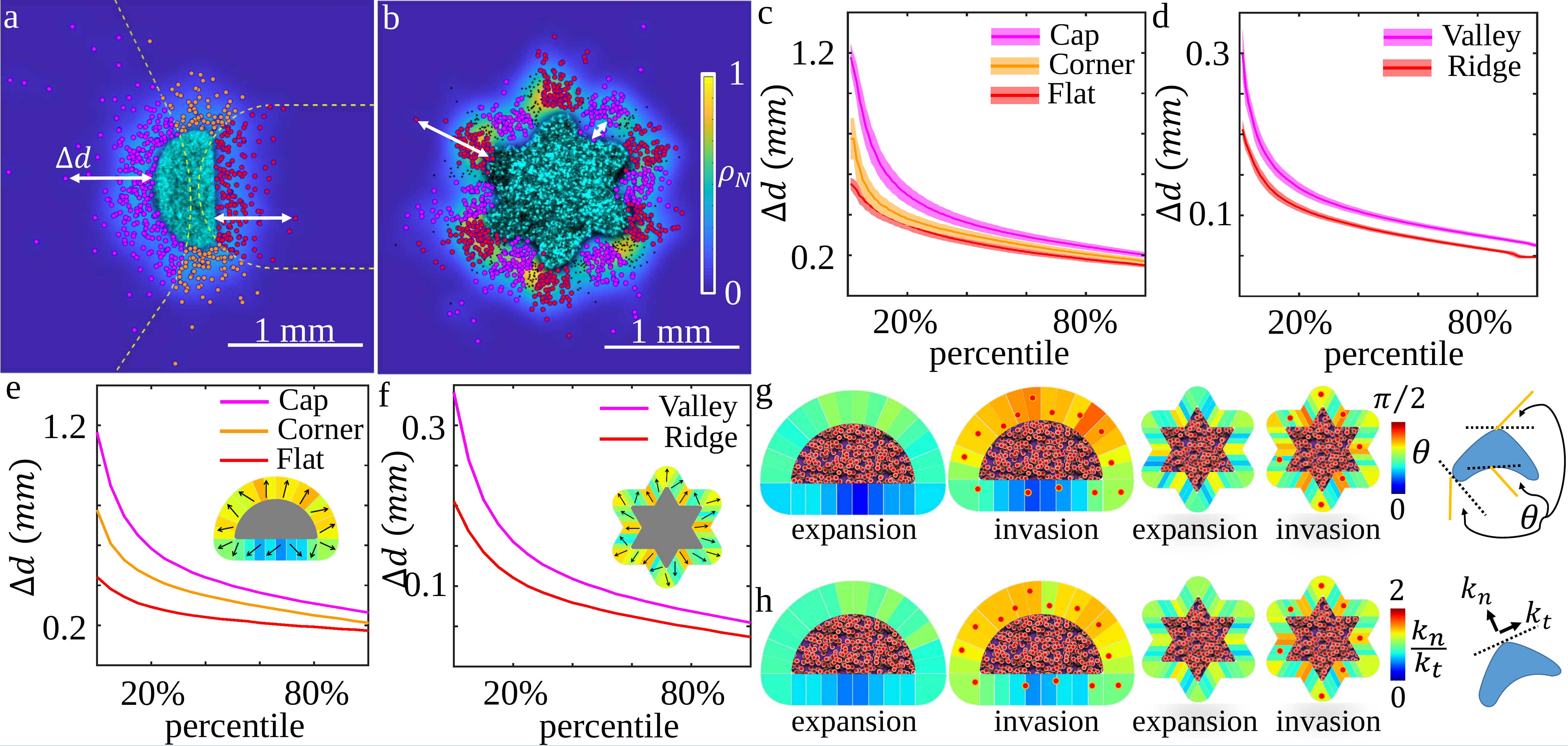}
\caption{Collective invasion of tumor organoids with complex
shapes. (a-b) Snapshots of semicircle and star-shaped diskoids at
day 10 and day 3 respectively. The images of the organoids at day zero are surrounded by the normalized cell density $\rho_N$ (see methods for definition). Dots represent locations of all the
disseminated cells experimentally observed for each tumor shape. In (a) the ECM space is divided into cap, corner and flat regions separated by the yellow dashed curves. In (b) the ECM space is divided into ridge and valley regions. Cells in
different regions are labeled with different colors. The invasion depth $\Delta d$ of a cell is the
distance between the cell and the boundary of the organoid at day
zero (white arrows show examples). (c-d) Ranked average invasion
depths (RAID) of semicircle and star-shaped diskoids. The solid
lines and shaded areas represent the means and standard deviations
obtained from 500 bootstraps. In (a-d) approximately 1000
disseminated cells are identified and included in the statistics
from N=4 (semicircle diskoid) and N = 5 (star diskoid) biological
replicas respectively. (e-f) Simulated RAID profiles for
semicircle and star shaped tumors. Insets: magnitude (heatmap, normalized by maximum), and direction (arrows) 
of average cell velocity near the tumor organoids. (g-h) The simulated average
fiber orientation and micromechanical anisotropy in a layer of $\sim 100 \mu m$ surrounding the
tumors after the expansion and invasion phases. Here $\theta$ is
defined as the acute angle between a fiber and the local
tangential direction of the tumor
boundary, while $k_n$ and $k_t$ are defined previously (Fig. S8).}\label{fig_complexgeometry}
\end{figure*}

{\bf Collective invasion of tumor organoids with complex shapes.}
Having tested our simulation model against experimental results for circular and triangular tumors, we ask if the mechanical principles considered in our model are sufficient to predict the invasiveness of tumors with
more complex geometries. We focus on two particular shapes of
tumors: semicircle and star. The boundary of a semicircle contains
regions of both positive and zero curvature, therefore can be
considered to be a hybrid of circular and triangular tumors. A
star shape, on the other hand, contains both convex and concave
surfaces.

Taking into account of the asymmetric shape of semicircle
diskoids, we divide the ECM space into cap, corner and flat
regions (Fig. \ref{fig_complexgeometry}a). We manually identify
all disseminated cells and their locations after 10 days from
seeding the original tumor. To help visualize the spatial
distribution of the cells we compute the normalized cell density
$\rho_N$ (Methods), which represents a dilution factor: if
$\rho_N=1$ then the local cell density is the same as in the
original tumor, where all cells are presumably uniformly
distributed in the original tumor.

We find the cell density is almost uniform surrounding the
original semicircle diskoid, and decreases rapidly at larger
distance. However, the flat region contains fewer
cells that migrate deeply into the ECM space from the diskoid
boundary. To further quantify the relation between invasiveness
and local geometry of tumors, we calculate the ranked average
invasion depths (RAID) $\bar{\Delta D} (f)$. In particular, we
first measure the invasion depth $\Delta d_i$ of each cell $i$ as
the distance of the cell from the original tumor boundary (arrows
in Fig. \ref{fig_complexgeometry}a,b). We then compute
$\bar{\Delta D} (f)$ as the average invasion depth of cells in the
top $f$ percentile ranked by $\Delta d_i$. Using the metric RAID,
we compare the invasiveness of cells in each of the three regions
of the ECM surrounding semicircle diskoids. As shown in Figure.
\ref{fig_complexgeometry}c, cells in the cap region are leading
the dissemination. For instance, the top 10\% invaders in the cap
region have an average invasion depth of 672 $\mu$m, while the top
10\% invaders in the corner and flat region have an average
invasion depth of 470 $\mu$m. At a percentile of 5\%, cells in the
cap region have a lead of 200 $\mu$m than cells in the flat
region.

Consistent with the previous observations in Fig.
\ref{fig_invasion_depth}, the invasiveness of semicircular
diskoids provides strong evidence that local geometry regulate
cancer cell dissemination. In particular, a positive curvature in
the tumor surface accelerates the overall invasion
\cite{Celeste2012_epithelialcurvature}.

We have also quantified the invasiveness of star-shaped diskoids
after 3 days of seeding the tumor. In particular, we divide the
ECM space into regions that are in the direction of the tips
(positive curvature), and regions that are in the direction of
valleys (negative curvature). Cells in the buffer regions (black
dots in Fig. \ref{fig_complexgeometry}b) are excluded from the
analysis.

By measuring RAID we find that overall cells in the valley region
possess larger invasion depth (Fig. \ref{fig_complexgeometry}d),
suggesting that negative curvature accelerates cell dissemination
even more than positive curvature. Of note, at 10 days, the disseminated cells become uniformly distributed in all directions. This is due to the proximity of the ridge and valley regions as well as the mixing caused by lateral motion of the cells,

These experimental results agree well with the predictions of our
simulations as shown in Fig. \ref{fig_complexgeometry}e and f.
Furthermore, our simulations also reveal the ECM remodeling by the
tumor organoids. Fig. \ref{fig_complexgeometry}(g-h) shows the
average fiber orientation and microscopic anisotropy in the expansion and
invasion phases. For semicircle
organoids, fibers in the flat region remain tangentially aligned
to the tumor boundary through the whole process; whereas fibers in
the cap region are re-oriented radially by cellular traction force
during the invasion phase. For star-shaped organoids, fiber
orientation in the ridge region rotates from tangential of the
tumor boundary to random alignment; whereas fibers in the valley
region are pulled normal to the tumor boundary during the invasion
phase. The structual anisotropy translates directly to the micromechanical anisotropy, such that the ECM is stiffer in the direction parallel to the fiber alignment. These results confirm that local geometry program
collective force generation and ECM remodeling by the cancer
cells, which modulates the rate of dissemination of the tumors.


In this letter, we demonstrate a previously unrecognized mechanism
of cell-cell interaction that coordinates multicellular dynamics.
This mechanism does not require direct contact between cells such
as cadherin-based adhesion
\cite{Ewald2013collectiveinvasion,Maheswaran2014} or contact
inhibition
\cite{Rappel2014_contactinhibition,Herbert2016contactinhibition},
nor it relies on the cooperation of leader-follower phenotypes
\cite{Friedl2007_individualtocollective,Sahai2007_fibroblastcancer,Haga2015,Wong2016_leadercell}.
Instead, we show both experimentally and computationally that
cells collectively apply forces to their ECM
\cite{Sun2017collectiveforce}, which in turn provides mechanical
cues to bias cell motility. Because collective force generation
can be controlled by geometry, we find dissemination of cancer
cells from tumor organoids are dependent on tumor geometry.

Our results provide physical insights for processes in cancer
biology and morphogenesis. Clinical studies have shown that
collagen fibers aligned tangentially and normally to tumor
boundary correspond to opposite prognosis
\cite{Keely2006,Keely2011}. While the origin of tumor-associated
ECM misalignment is unclear, our results suggest that tumor
geometry is an important contributing factor. On the other hand,
mesenchymal cell migration is often considered as an single-cell
process during development and diseases
\cite{Erik2006_EMT,Nieto2009}. Our model system show that
underlying multicellular coordination may take place in the form
of collective force generation and ECM remodeling. Together, we
find that 3D collective cell migration may exploit the mechanical
feedback between force-generating cells and reconfigurable ECM as
an indirect yet effective channel of communication .

Finally, our results show that 3D migrating cells represent a
distinct class of active particles which actively re-sculpture
their microenvironment and respond to the cues generated by
themselves and others. Future research is needed to systematically
investigate the collective dynamics of such active particles as a
route to understand general living systems.

\subsection*{Methods}

\textit{Sample preparation.} See Supplementray Information for
details.

\textit{Image analysis.} Confocal images were taken using Leica
SPE at a rate of 30 min per frame for continuous imaging or at
days 0, 1, 3, 5, 10 for discrete imaging. Cell locations were
projected onto the $x-y$ plane, whereas movement in the $z$
direction is relatively small \cite{Sun2016DIGME}. The invasion depths were
manually measured with the help of NIH ImageJ. The deformation of
the ECM was measure using PIVlab implemented on Matlab. To
calculate the velocity field in the ECM we perform PIV analysis on
image pairs with 2-hour delay.

To approximate the cell density from the scattered cell locations we use a gaussian kernel:
\begin{eqnarray}
\rho_N (\mathbf{r}) = \frac{A_0}{m}\sum_{i=1}^m \frac{1}{2\pi\sigma^2}e^{-\frac{(\mathbf{r}-\mathbf{r}_i)^2}{2\sigma^2}}
\end{eqnarray}
Here we choose the kernel width $\sigma$ to be 80 $\mu$m,
approximately twice the size of a cell. $m$ is the total number of
disseminated cells, and $A_0$ is the area of the original tumor
diskoid. To calculate the ranked average invasion depth (RAID) we
used the Matlab \texttt{quantile()} function to select the data
for averaging.

\textit{Computation.} See supplementary information for details.

\subsection*{Data Availability}
All data and computer codes are available from the authors upon reasonable request.
\subsection*{Acknowledgment}
The research is supported by a Scialog Program sponsored jointly by Research Corporation for Science Advancement and the Gordon and Betty Moore Foundation through a grant to Oregon State University by the Gordon and Betty Moore Foundation. B. S. is partially supported by the Medical Research Foundation of Oregon and SciRIS-II award from Oregon State University. J. K. is partially supported by the National Science Foundation grant PHY-1400968.

\subsection*{Author Contributions}
B. S. and Y. J. designed the research and oversaw the experimental and computational studies respectively. J. K., Y. Z., A. A., H. N., J. T. collected data. All authors analyzed data and wrote the manuscript. Y. Z. and H. N. thank Arizona State University for the University Graduate Fellowship.





\bibliography{CollectiveInvasion.bib}

\providecommand{\noopsort}[1]{}\providecommand{\singleletter}[1]{#1}%
\begin{thebibliography}{10}

\bibitem{Segall_collective_review}
P.~Friedl, J.~Locker, E.~Sahai, and J.~E. Segall.
\newblock Classifying collective cancer cell invasion.
\newblock {\em Nat. Cell Biol.}, 14(7):777--783, 2012.

\bibitem{Pascal2017_collectivereview}
Vincent Hakim and Pascal Silberzan.
\newblock Collective cell migration: a physics perspective.
\newblock {\em Reports on Progress in Physics}, 80(7):076601, 2017.

\bibitem{Ewald2013collectiveinvasion}
KJ~Cheung, E~Gabrielson, Z~Werb, and AJ~Ewald.
\newblock Collective invasion in breast cancer requires a conserved basal
  epithelial program.
\newblock {\em Cell}, 155(7):1639, 2013.

\bibitem{Maheswaran2014}
Nicola Aceto, Aditya Bardia, David.T. Miyamoto, Maria.C. Donaldson, Ben.S.
  Wittner, Joel.A. Spencer, Min Yu, Adam Pely, Amanda Engstrom, Huili Zhu,
  Brian.W. Brannigan, Ravi Kapur, Shannon.L. Stott, Toshi Shioda, Sridhar
  Ramaswamy, David.T. Ting, Charles.P. Lin, Mehmet Toner, Daniel.A. Haber, and
  Shyamala Maheswaran.
\newblock {Circulating Tumor Cell Clusters Are Oligoclonal Precursors of Breast
  Cancer Metastasis}.
\newblock {\em Cell}, 158(5):1110--1122, 2014.

\bibitem{Friedl2007_individualtocollective}
Katarina Wolf, Yi~I. Wu, Yueying Liu, J{\"{o}}rg Geiger, Eric Tam, Christopher
  Overall, M.~Sharon Stack, and Peter Friedl.
\newblock {Multi-step pericellular proteolysis controls the transition from
  individual to collective cancer cell invasion}.
\newblock {\em Nature Cell Biology}, 9(8):893--904, 2007.

\bibitem{Sahai2007_fibroblastcancer}
Cedric Gaggioli, Steven Hooper, Cristina Hidalgo-Carcedo, Robert Grosse,
  John~F. Marshall, Kevin Harrington, and Erik Sahai.
\newblock {Fibroblast-led collective invasion of carcinoma cells with differing
  roles for RhoGTPases in leading and following cells}.
\newblock {\em Nature Cell Biology}, 9(12):1392--1400, 2007.

\bibitem{Marcus2017}
J~Konen, E~Summerbell, B~Dwivedi, K~Galior, Y~Hou, L~Rusnak, A~Chen, J~Saltz,
  W~Zhou, L~H Boise, P~Vertino, L~Cooper, K~Salaita, J~Kowalski, and A~I
  Marcus.
\newblock {Image-guided genomics of phenotypically heterogeneous populations
  reveals vascular signalling during symbiotic collective cancer invasion}.
\newblock {\em Nat. Comm.}, 8:15078, 2017.

\bibitem{Sun2016DIGME}
A.~A. Alobaidi and B.~Sun.
\newblock Probing three-dimensional collective cancer invasion with digme.
\newblock {\em Cancer Convergence}, accepted, 2017.

\bibitem{Sun2017ECMplasticity}
J.~Kim, J.~Feng, C.~A.~R. Jones, X.~Mao, L.~M. Sander, H.~Levine, and B.~Sun.
\newblock Stress-induced plasticity of dynamic biopolymer networks.
\newblock {\em Nature Communications}, 2017, in press.

\bibitem{Sun2014gel}
C.~A. Jones, L.~Liang, D.~Lin, Y.~Jiao, and B.~Sun.
\newblock The spatial-temporal characteristics of type i collagen-based
  extracellular matrix.
\newblock {\em Soft Matter}, 10(44):8855--8863, 2014.

\bibitem{Fabry2016_3dtractionforce}
J~Steinwachs, C~Metzner, K~Skodzek, N.Lang, I~Thievessen, C~Mark,
  S~M{\"{u}}nster, K~E Aifantis, and B~Fabry.
\newblock {Three-dimensional force microscopy of cells in biopolymer networks}.
\newblock {\em Nat. Methods}, 13:171, 2016.

\bibitem{han2016oriented}
Weijing Han, Shaohua Chen, Wei Yuan, Qihui Fan, Jianxiang Tian, Xiaochen Wang,
  Longqing Chen, Xixiang Zhang, Weili Wei, Ruchuan Liu, et~al.
\newblock Oriented collagen fibers direct tumor cell intravasation.
\newblock {\em Proceedings of the National Academy of Sciences},
  113(40):11208--11213, 2016.

\bibitem{Levine2018durotaxis}
Jingchen Feng, Herbert Levine, Xiaoming Mao, and Leonard~M Sander.
\newblock Stiffness sensing and cell motility: Durotaxis and contact guidance.
\newblock {\em bioRxiv 320705}, 2018.

\bibitem{Robinson2013_mechanosensing}
Tianzhi Luo, Krithika Mohan, Pablo~A. Iglesias, and Douglas~N. Robinson.
\newblock {Molecular mechanisms of cellular mechanosensing}.
\newblock {\em Nature Materials}, 12:1064–1071, 2013.

\bibitem{Celeste2012_epithelialcurvature}
Eline Boghaert, Jason~P Gleghorn, KangAe Lee, Nikolce Gjorevski, Derek~C
  Radisky, and Celeste~M Nelson.
\newblock {Host epithelial geometry regulates breast cancer cell invasiveness.}
\newblock {\em Proceedings of the National Academy of Sciences},
  109(48):19632--7, 2012.

\bibitem{Rappel2014_contactinhibition}
Brian~A Camley, Yunsong Zhang, Yanxiang Zhao, Bo~Li, Eshel Ben-Jacob, Herbert
  Levine, and Wouter-Jan Rappel.
\newblock {Polarity mechanisms such as contact inhibition of locomotion
  regulate persistent rotational motion of mammalian cells on micropatterns.}
\newblock {\em Proceedings of the National Academy of Sciences},
  111(41):14770--5, 2014.

\bibitem{Herbert2016contactinhibition}
Juliane Zimmermann, Brian~A Camley, Wouter-Jan Rappel, and Herbert Levine.
\newblock {Contact inhibition of locomotion determines cell-cell and
  cell-substrate forces in tissues.}
\newblock {\em Proceedings of the National Academy of Sciences},
  113(10):2660--5, 2016.

\bibitem{Haga2015}
Naoya Yamaguchi, Takeomi Mizutani, Kazushige Kawabata, and Hisashi Haga.
\newblock {Leader cells regulate collective cell migration via Rac activation
  in the downstream signaling of integrin $\beta$1 and PI3K}.
\newblock {\em Scientific Reports}, 5:1--8, 2015.

\bibitem{Wong2016_leadercell}
Zachary~S. Dean, Paul Elias, Nima Jamilpour, Urs Utzinger, and Pak~Kin Wong.
\newblock {Probing 3D collective cancer invasion using double-stranded locked
  nucleic acid biosensors}.
\newblock {\em Analytical Chemistry}, 88(17):8902--8907, 2016.

\bibitem{Sun2017collectiveforce}
Amani~A Alobaidi, Yaopengxiao Xu, Shaohua Chen, Yang Jiao, and Bo~Sun.
\newblock {Probing cooperative force generation in collective cancer invasion}.
\newblock {\em Physical Biology}, 14(4):045005, jun 2017.

\bibitem{Keely2006}
Paolo~P Provenzano, Kevin~W Eliceiri, Jay~M Campbell, David~R Inman, John~G
  White, and Patricia~J Keely.
\newblock {Collagen reorganization at the tumor-stromal interface facilitates
  local invasion}.
\newblock {\em BMC Medicine}, 4(1):38, 2006.

\bibitem{Keely2011}
M~W Conklin, J~C Eickhoff, K~M Riching, C~A Pehlke, K~W Eliceiri, P~P
  Provenzano, A~Friedl, and P~J Keely.
\newblock {Aligned Collagen Is a Prognostic Signature for Survival in Human
  Breast Carcinoma}.
\newblock {\em Am. J. Path.}, 178(3):1221, 2011.

\bibitem{Erik2006_EMT}
Jonathan~M Lee, Shoukat Dedhar, Raghu Kalluri, and Erik~W Thompson.
\newblock {The epithelial-mesenchymal transition: new insights in signaling,
  development, and disease.}
\newblock {\em The Journal of cell biology}, 172(7):973--81, 2006.

\bibitem{Nieto2009}
Herv{\'{e}} Acloque, Meghan~S Adams, Katherine Fishwick, Marianne
  Bronner-Fraser, and M~Angela Nieto.
\newblock {Epithelial-mesenchymal transitions: the importance of changing cell
  state in development and disease.}
\newblock {\em The Journal of clinical investigation}, 119(6):1438--49, 2009.

\end{thebibliography}
\bibliographystyle{unsrt}

\end{document}